\documentclass[english,5p,twocolumn,times]{elsarticle}
\usepackage[T1]{fontenc}
\usepackage[utf8]{inputenc}
\usepackage{geometry}
\usepackage{babel}
\usepackage{array}
\usepackage{verbatim}
\usepackage{multirow}
\usepackage{amsbsy}
\usepackage{graphicx}
\usepackage{rotating}
\usepackage[unicode=true]
 {hyperref}

\makeatletter

\providecommand{\tabularnewline}{\\}

\usepackage{color}
\usepackage{graphics}
\usepackage{array}
\usepackage{graphicx}
\usepackage{lastpage}
\usepackage{float}

\usepackage{times}

\journal{XXX}

\usepackage{amsmath,amssymb}

\usepackage{colortbl}
\usepackage{multicol}

\definecolor{v1}{rgb}{0.0,0.8,0.0}
\definecolor{v2}{rgb}{0.8,1.0,0.0}
\definecolor{v3}{rgb}{0.8,0.8,0.8}
\definecolor{v4}{rgb}{0.9,0.5,0.0}
\definecolor{v5}{rgb}{0.8,0.0,0.0}

\usepackage{enumitem}
\setlist{nosep} % or \setlist{noitemsep} to leave space around whole list

\makeatother

\begin{document}
\sloppy
\begin{frontmatter}

\title{Impact on clinical guideline adherence of Orient-COVID,\\
a CDSS based on dynamic medical decision trees for COVID19 management:\\
a randomized simulation trial}

\author[label1]{Mouin Jammal}
\ead{mouinjammal@yahoo.fr}

\author[label2,label7]{Antoine Saab}
\ead{antoine_saab@outlook.com}

\author[label3,label7]{Cynthia Abi Khalil}
\ead{Cynthiakjammal@outlook.com}

\author[label4]{Charbel Mourad}
\ead{charbel.j.mourad@hotmail.com}

\author[label5,label6]{Rosy Tsopra}
\ead{rosytsopra@gmail.com}

\author[label2]{Melody Saikali}
\ead{melodysaikali@hotmail.com}

\author[label7]{Jean-Baptiste Lamy\corref{cor1}}
\ead{jean-baptiste.lamy@inserm.fr}
\cortext[cor1]{Corresponding author}

\address[label1]{Department of Internal Medicine, Lebanese Hospital Geitaoui-UMC, Beirut, Lebanon}

\address[label2]{Quality and Patient Safety Department, Lebanese Hospital Geitaoui-UMC, Beirut, Lebanon}

\address[label3]{Nursing Administration, Lebanese Hospital Geitaoui-UMC, Beirut, Lebanon}

\address[label4]{Department of Medical Imaging, Lebanese Hospital Geitaoui-UMC, Beirut, Lebanon}

\address[label5]{Université Paris Cité, Sorbonne Université, Inserm, Centre de Recherche des Cordeliers, F-75006 Paris}
\address[label6]{Department of Medical Informatics, AP-HP, Hôpital Européen Georges-Pompidou, F-75015 Paris, France}

\address[label7]{INSERM, Université Sorbonne Paris Nord, Sorbonne Université, Laboratory of Medical Informatics and Knowledge Engineering in e-Health (LIMICS), Paris, France}
\begin{abstract}
\noindent \textbf{Background:} The adherence of clinicians to clinical
practice guidelines is known to be low, including for the management
of COVID-19, due to their difficult use at the point of care and their
complexity. Clinical decision support systems have been proposed to
implement guidelines and improve adherence. One approach is to permit
the navigation inside the recommendations, presented as a decision
tree, but the size of the tree often limits this approach and may
cause erroneous navigation, especially when it does not fit in a single
screen.

\noindent \textbf{Methods:} We proposed an innovative visual interface
to allow clinicians easily navigating inside decision trees for the
management of COVID-19 patients. It associates a multi-path tree model
with the use of the fisheye visual technique, allowing the visualization
of large decision trees in a single screen. To evaluate the impact
of this tool on guideline adherence, we conducted a randomized controlled
trial in a near-real simulation setting, comparing the decisions taken
by medical students using Orient-COVID with those taken with paper
guidelines or without guidance, when performing on six realistic clinical
cases.

\noindent \textbf{Results:} The results show that paper guidelines
had no impact (p=0.97), while Orient-COVID significantly improved
the guideline adherence compared to both other groups (p<0.0003).
A significant impact of Orient-COVID was identified on several key
points during the management of COVID-19: ordering troponin lab tests,
prescribing anticoagulant and oxygen therapy. A multifactor analysis
showed no difference between male and female participants.

\noindent \textbf{Conclusions:} The use of an interactive decision
tree for the management of COVID-19 significantly improved the clinician
adherence to guidelines. Future works will focus on the integration
of the system to electronic health records and on the adaptation of
the system to other clinical conditions.
\end{abstract}
\end{frontmatter}

\section{Introduction}

The US Institute of Medicine's influential report ``To Err Is Human''
\citep{visu-cogn-erreur2} created awareness that medical error is
a major cause of avoidable mortality, morbidity and inappropriate
use of resources. With the increasing recognition of shortcomings
of healthcare systems, practice guidelines were widely advocated as
a means of encouraging compliance with evidence-based practice, leading
to the ``guidelines movement'' \citep{Fox2009}. Clinical Practice
Guidelines (CPGs) are text documents summarizing recommended practices
for a specific condition, with the rationale and supporting evidence.
CPGs may include flowchart clinical algorithms.

There is evidence that CPGs can improve clinical outcomes, but also
that the level of adherence is low in practice \citep{Sharmin2021}.
Paper guidelines provide limited support to clinicians for finding
patient-specific recommendations \citep{Kilsdonk2016}. The adherence
to CPGs is impaired by many factors \citep{Lugtenberg2011,Sinuff2007,Shaneyfelt1999,Shaughnessy2016,Sultan2023}
including: (1) inaccessibility of guidance at the point of care: CPGs
are long documents, difficult to read during medical consultations,
(2) difficulties of application to local settings, (3) oversimplification:
most CPGs address a single disease while many patients have multiple
comorbidities, (4) ambiguity: guidelines are not written in a formal
language, and (5) lack of integration of patient values and goals.

Computerized Decision Support Systems (CDSSs) are designed to implement
CPGs and help clinicians make decision about individual \citep{berner2007,Hak2022,Souza-Pereira2020}.
Evidence suggests that CDSSs can positively impact care processes
\citep{Kwan2020} and guideline adherence \citep{Gholamzadeh2023}.
CDSSs can mitigate to a considerable degree the criticisms frequently
made about CPGs.

The first step in CDSS design is to formalize the medical knowledge
that is informally described in CPGs, using a variety of computer-interpretable
formats. The second step is to develop a computer application that
presents the knowledge conveniently, \emph{e.g.} by triggering alerts
\citep{jiba-asti-critique} or allowing interactive navigation in
the recommendations, presented as a tree or as a sequence of questions
\citep{aidedecision-asti-modeguide}. However, trees are often too
large to be presented in their entirety on the screen, the navigation
is then laborious and can be a significant cause of errors (up to
44\% error in complex situations \citep{aidedecision-asti-guide-gbp-evalenligne}).

The present study is part of a larger project, Orient-COVID, aimed
at designing a CDSS to support the management of patients with COVID19.
The CDSS relies on an innovative visual interface for navigating in
a decision tree. The present work aims to measure the impact of Orient-COVID
on physician adherence, through a randomized controlled trial methodology
in a simulated setting, versus paper guidelines, and versus the absence
of support. The paper will follow the amendments to the STROBE guidelines
for simulation-based research \citep{Cheng2016}.

\begin{figure*}[!p]
\begin{centering}
\includegraphics[width=1\textwidth]{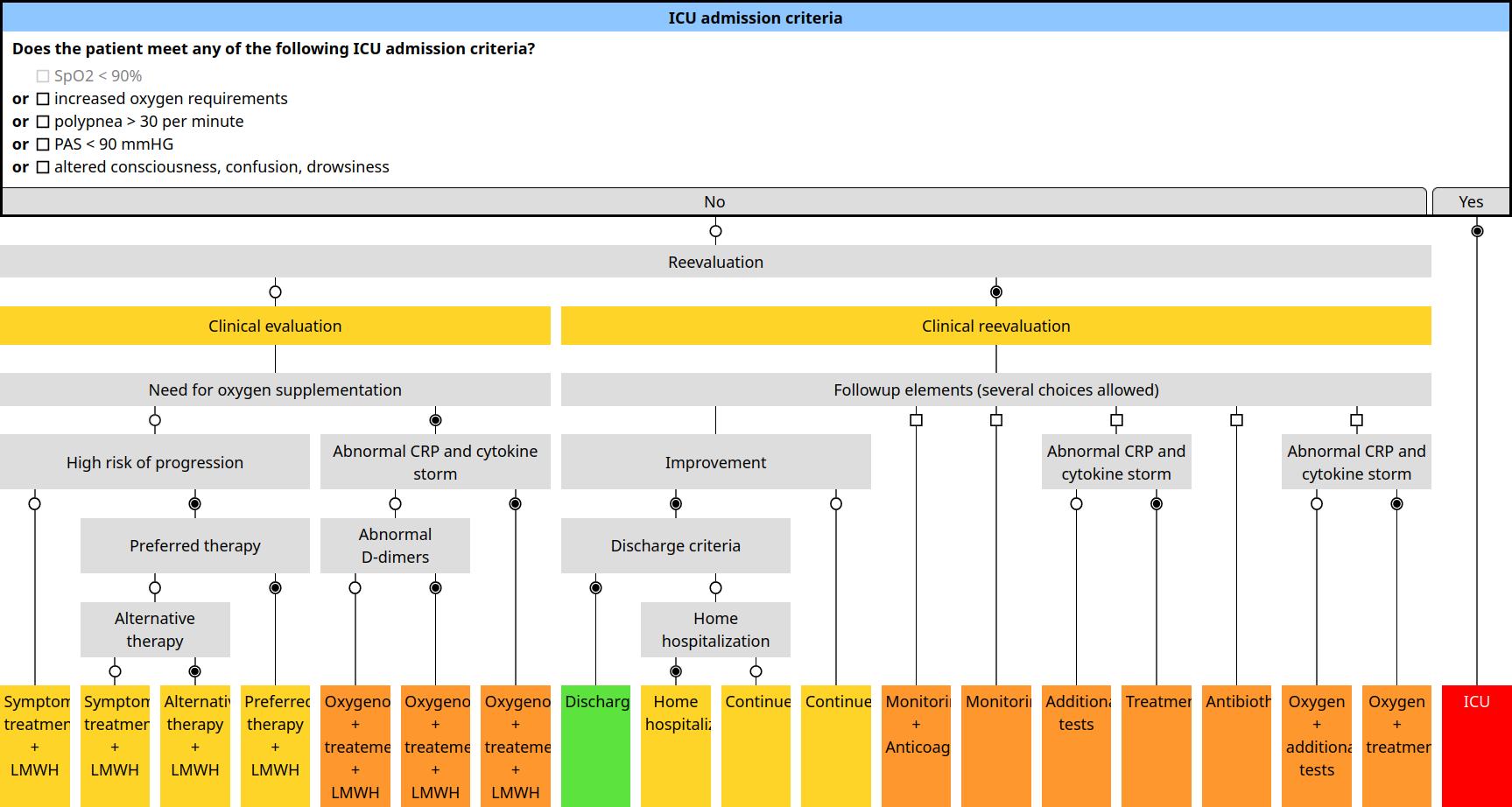}
\par\end{centering}
\caption{\label{fig:arbre_1}Screenshot of the interactive decision tree for
the management of hospitalized Covid-19 patients, before any user
interaction. It gives an overview of the entire decision process,
at a glance. Most nodes are yes/no questions, and use checked/unchecked
radio buttons as symbols on the edge. To interact with the tree, the
user can either click on the button at the bottom of a current node
(\emph{e.g.} “Yes” or “No”), or directly click on any node, for performing
a faster or backward navigation.}

\bigskip{}

\bigskip{}

\bigskip{}

\begin{centering}
\includegraphics[width=1\textwidth]{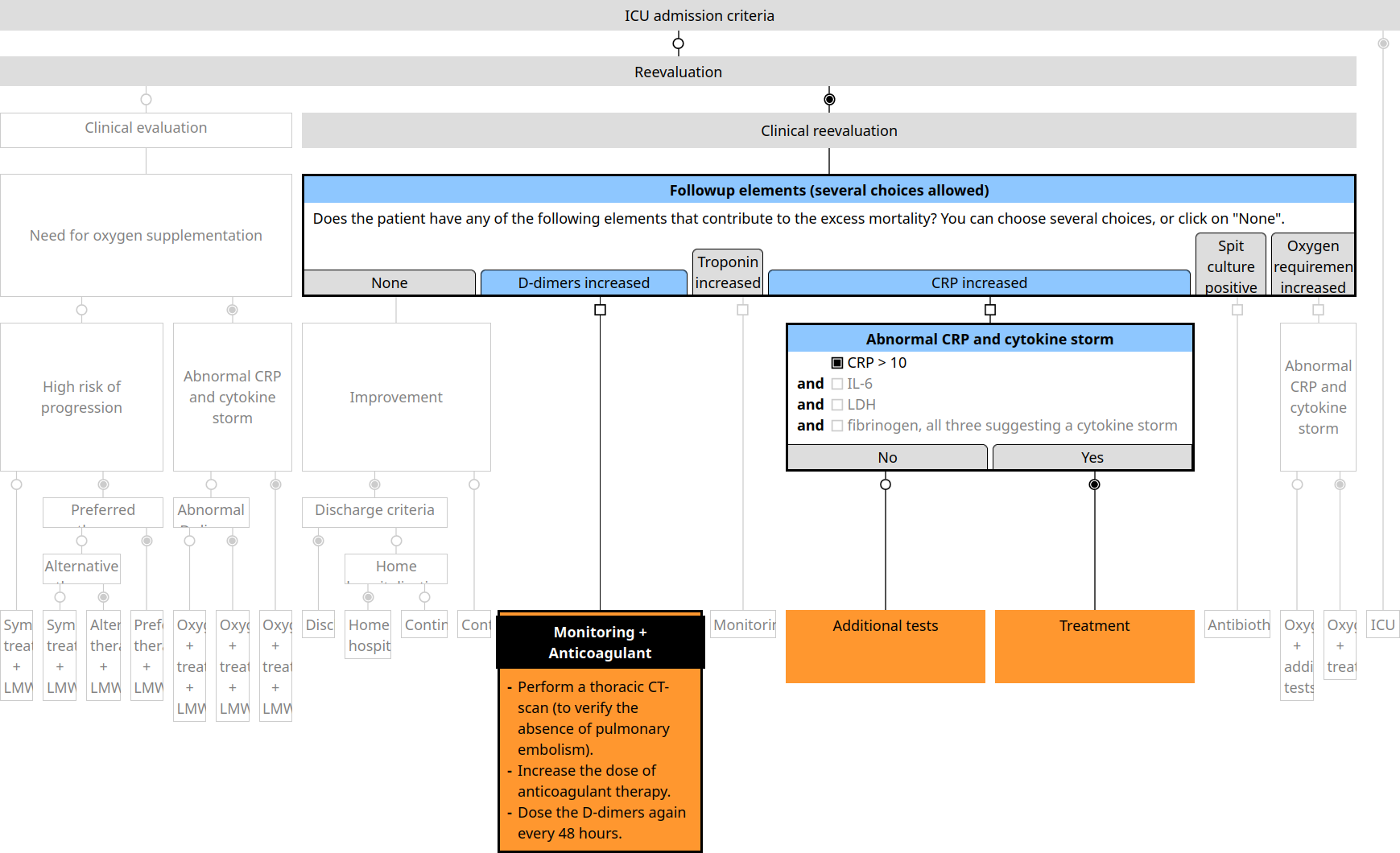}
\par\end{centering}
\caption{\label{fig:arbre_2}Screenshot of the interactive multi-path decision
tree for the management of hospitalized Covid-19 patients, after some
user interactions. Three nodes are current: two question nodes (labeled
“Followup elements”, a multiple-choice question node, and “Abnormal
CRP and cytokine storm”) and a recommendation node (“Monitoring +
Anticoagulant”). Two other recommendations are still accessible for
future navigation (“Additional tests” and “Treatment”). Notice the
parts of the tree that have not been selected have been grayed and
squeezed, thanks to fisheye.}
\end{figure*}

\begin{figure}
\begin{centering}
\includegraphics[width=0.8\columnwidth]{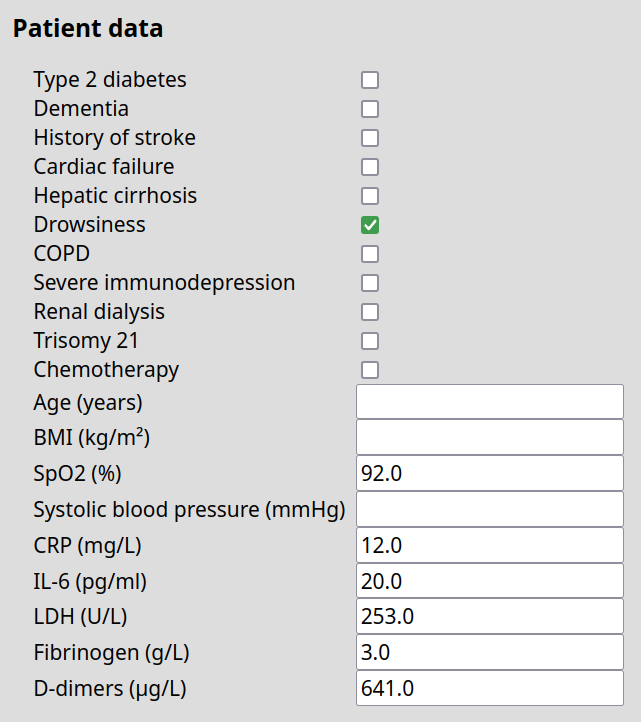}
\par\end{centering}
\caption{\label{fig:Form}Screenshot of the patient data entry form.}
\end{figure}

\section{\label{sec:Methods}Materials and methods}

\subsection{Brief description of Orient-COVID}

Orient-COVID is based on medical decision trees established from a
review of international best practice guidelines for the management
of COVID19 and formalized by a multidisciplinary team, including doctors,
nurses and specialists in medical informatics. The decision trees
have been structured in a formal ontology, and stored in an RDF quadstore.

The CDSSs proposes an interactive navigation through the decision
trees. It includes two innovative features for reducing the size of
the tree and permitting its presentation in its entirety on the screen:
(1) the use of the fisheye technique reduces the space devoted to
the unselected parts of the tree, and (2) the use of a multi-path
decision tree model \citep{Guo2013} allows the user selecting several
paths at specific nodes. This is particularly useful when the CPG
considers several risk factors or followup elements, and proposes
a specific independent response for each. In such situations, the
multi-path tree model avoid duplication of parts of the tree. Instead,
several paths are selected and each leads to a distinct recommendation.

Orient-COVID was developed as a client-server web application in Python
using Brython, a Javascript-compiled version of Python and Owlready,
a module for ontology-oriented programming \citep{Lamy2017_5}. The
role of the server is limited and most of the program is implemented
in the client. This allows patient data to remain on the client and
thus supports data privacy.

Figure \ref{fig:arbre_1} shows the multi-path decision tree for hospitalization
of Covid-19 patients, before user navigation, and Figure \ref{fig:arbre_2}
during user interaction. Orient-COVID also proposes a patient data
entry form, where the clinician can optionally enter patient data
(Figure \ref{fig:Form}). The data is used for triggering a personalized
semi-automatic navigation in the decision tree, hence accelerating
the navigation.

For more details on Orient-COVID, please refer to \citep{Lamy2023}
and to the demonstration website:

\href{http://www.lesfleursdunormal.fr/appliweb/orient_covid}{http://www.lesfleursdunormal.fr/appliweb/orient\_covid}.

\subsection{Study design}

This is a single-center, 3-arm parallel group unblinded randomized
controlled study performed at the Lebanese Hospital Geitaoui, a 250-beds
University Medical Center in Beirut, Lebanon. The study was performed
between December 2023 and February 2024. The participants performed
on clinical cases, and no real patients were involved.

The protocol and informed consent documents were reviewed and approved
by the hospital institutional review board.

\begin{comment}
The patient data that is necessary for elaborating the clinical cases
were anonymized before the experimental steps. Members of the expert
panel which prepared the cases signed a nondisclosure agreement relative
to the patient data that was used at this step.
\end{comment}

\subsection{Recruitment}

Participants were medical students and residents on rotation at the
Lebanese Hospital Geitaoui-UMC. Enrollment was open after a communication
about the study through diffusion lists. The participants were equally
randomized in 3 groups: group A (no guidance), group B (paper guidance)
and group C (Orient-COVID). Participants were remunerated for their
participation in the study.

Upon their enrollment, participants received an information notice
about the study method and protocol.%
\begin{comment}
A number of medical residents initially enrolled were excluded because
they stated their will to withdraw, but were later replaced by randomly
picked medical students and residents from the pool of initially enrolled,
in the same year of studies.
\end{comment}
{} Each participant performed sequentially all six clinical cases in
the presence of the same senior medical professional (internal medicine
physician with more than 10 years of postgraduate clinical experience,
and thorough experience in managing COVID19 cases). His role was to
perform the simulation in total neutrality, including asking the participant
to state his decision at all steps, and recording the participant
answers.

\subsection{Clinical cases and gold standard}

Six COVID19 clinical cases were created by a panel of medical experts,
inspired by retrospective anonymized data of real patients admitted
to the hospital between January and December 2022. This ensured near-to-real
patient data for the simulation. The cases covered a number of common
COVID19 hospital scenarios in terms of severity, with different outcomes
(healed, deceased, transferred to higher level of care). For each
case, experts defined through consensus and in accordance with CPGs
a set of time-dependent (upon admission, post-24h, upon discharge)
diagnostic, clinical and therapeutic decisions, and then analyzed
the patient medical file to verify patient clinical pathways, outcomes
and conformity with the CPGs. These decisions relative to each case
constituted the gold standard for the study. Cases are labeled thereafter
1, 2, 3, 4, 5, 6 (Supplementary Material \#1).

\subsection{Protocol}

The participant signed an informed consent and disclosed demographic
data (year of residency, year of birth, sex, university grade in the
past academic year, prior experience with management of COVID19 hospital
cases, estimated number of COVID19 cases managed, date of last COVID19
case managed). The participant was then given an appointment for the
simulation session, which comprised the following steps: 
\begin{enumerate}
\item The participant received an information notice about the study objectives,
methods and steps.
\item If the participant was in Group B, he was presented with the paper
CPG.
\item If the participant was in Group C, he was presented with Orient-COVID.
\item The senior evaluator presented the six cases sequentially to the participant,
instructed the participant to formulate a decision according to the
predefined decision checklist. The participant decisions were recorded
by the evaluator and no critical feedback was given to the participant.
\end{enumerate}

\subsection{\label{subsec:Data-collected}Data collected}

For each clinical case, 22 decision criteria were considered, including:
\begin{itemize}
\item 9 criteria regarding the initial evaluation, and consisting of examinations
or lab tests that can be ordered: EKG (electrocardiogram), chest CT
(computed tomography), general blood test, CRP (C-reactive protein),
LDH (lactate dehydrogenase), troponin, D-Dimers, ferritin, Il 6 (interleukin
6).
\item 6 criteria regarding the initial decision, consisting of 2 criteria
relative to the decision to hospitalize the patient and the level
of care (\emph{e.g.} ICU or not), and 4 criteria about drug prescriptions:
antibiotics, steroids, anticoagulant, oxygen.
\item 7 criteria regarding the reevaluation of the patient, mixing medical
assessment of the patient status, examination and lab test ordering:
clinical status, oxygen need, fever, blood test, chest CT, reevaluation
decision, plan of care.
\end{itemize}
For each case solved by a participant, all criteria were assessed
and counted for 1 point. Whenever the participant’s answer was in
concordance with the gold standard for the criteria, 1 point was awarded.
When the participant answer was incorrect, 0 point was awarded. The
total score, with a maximum of 22, was the primary outcome. Secondary
outcomes were each criterion considered individually, and impact of
the various factors (age, sex,...).

\subsection{Number of subjects}

Based on a power of 0.8, a risk $\alpha=5\%$, a mean difference in
score of 2 points, and a standard deviation of 4 points, the minimum
number of clinical case solved per group is 60, leading to 10 participants
by group.

\subsection{Statistical analysis}

Statistical analysis was performed using R software, with a risk $\alpha=5\%$
and bilateral tests. For the analysis, the base unit is the clinical
case solved by a participant.

First, the mean score obtained in the three groups was compared with
Welch Two Sample t-test, in two-by-two comparisons (3 tests).

Second, the impact of Orient-COVID was tested on each of the 22 criteria
using Welch Two Sample t-test. Due to the relatively important number
of criteria, we applied Bonferroni correction: we considered a significance
threshold $\alpha'=\frac{0.05}{22}=0.0023$.

Third, in addition to the use of Orient-COVID, the following factors
were tested: participant sex, age, grades obtained at university,
number of Covid-19 cases treated in the last year by the participant,
and clinical case ID. For each factor, we performed a linear mixed
model (LMM) analysis on the score, considering two fixed-effect factors:
Orient-COVID and the factor to test (including the potential interaction
between them). The participant ID was added as a random-effect factor.
\begin{comment}
LMM was performed using the ``nlme'' R package, and then the ``car''
package was used for computing type-III ANOVA.
\end{comment}

Supplementary Material \#2 include the dataset and \#3 the R sources.

\section{\label{sec:Results}Results}

\begin{table}[t]
\begin{centering}
\begin{tabular}{>{\raggedright}p{2.5cm}>{\raggedright}p{1.1cm}lc}
\textbf{Characteristic} & \textbf{Type} & \multicolumn{2}{c}{\textbf{Modalities / Aggregation}}\tabularnewline
\hline 
\multirow{2}{2.5cm}{Sex} & \multirow{2}{1.1cm}{nominal} & Male & 12 (40\%)\tabularnewline
 &  & Female & 18 (60\%)\tabularnewline
\hline 
 & \multirow{3}{1.1cm}{integer} & Mean & 25.9\tabularnewline
Age (years) &  & Min & 24\tabularnewline
 &  & Max & 33\tabularnewline
\hline 
\multirow{4}{2.5cm}{Study year} & \multirow{4}{1.1cm}{integer} & 7\textsuperscript{th} & 10\tabularnewline
 &  & 8\textsuperscript{th} & 9\tabularnewline
 &  & 9\textsuperscript{th} & 6\tabularnewline
 &  & 10\textsuperscript{th}+ & 5\tabularnewline
\hline 
\multirow{3}{2.5cm}{University grade: average grade in the previous year} & \multirow{3}{1.1cm}{integer (0-20)} & Mean & 15.3\tabularnewline
 &  & Min & 14\tabularnewline
 &  & Max & 17\tabularnewline
\hline 
\multirow{4}{2.5cm}{Number of Covid-19 cases treated} & \multirow{4}{1.1cm}{ordered nominal} & < 5 & 0\tabularnewline
 &  & 5 - 10 & 4 (13.3\%)\tabularnewline
 &  & 11 - 30 & 11 (36.7\%)\tabularnewline
 &  & > 30 & 15 (50\%)\tabularnewline
\hline 
\multirow{3}{2.5cm}{Group} & \multirow{3}{1.1cm}{nominal} & A & 10 (33.3\%)\tabularnewline
 &  & B & 10 (33.3\%)\tabularnewline
 &  & C & 10 (33.3\%)\tabularnewline
\hline 
\end{tabular}
\par\end{centering}
\begin{centering}
\begin{comment}
\begin{center}
\begin{tabular}{ccccccc}
 & \multicolumn{2}{c}{\% problems identified} & \multicolumn{2}{c}{\% interventions appropriate} & \multicolumn{2}{c}{mean / median time (minutes)}\tabularnewline
Case & with B & with A & with B & with A & with B & with A\tabularnewline
\hline 
A & 42.2\% & 77.4\% & 28.6\% & 64.1\% & 30.5 / 28.3 & 35.0 / 21.9\tabularnewline
B & 43.8\% & 84.4\% & 34.9\% & 80.7\% & 22.9 / 24.7 & 41.4 / 25.1\tabularnewline
C & 43.5\% & 67.9\% & 38.9\% & 66.3\% & 25.1 / 24.1 & 16.0 / 15.1\tabularnewline
D & 45.6\% & 59.5\% & 40.6\% & 49.6\% & 31.0 / 22.5 & 27.4 / 23.9\tabularnewline
\hline 
\textbf{Overall} & \textbf{43.5\%} & \textbf{71.9\%} & \textbf{34.8\%} & \textbf{64.7\%} & \textbf{27.3 / 25.4} & \textbf{29.3 / 20.2}\tabularnewline
 & \multicolumn{2}{c}{$p<2.2\times10^{-16}$ \textbf{({*})}} & \multicolumn{2}{c}{$p<2.2\times10^{-16}$ \textbf{({*})}} & \multicolumn{2}{c}{$p=0.56$}\tabularnewline
\hline 
\end{tabular}
\par\end{center}
\end{comment}
\par\end{centering}
\caption{\label{tab:demograph}Demographic characteristics of the recruited
participants.}
\end{table}

\subsection{Recruited participants}

Thirty participants were recruited in the study, 10 were allocated
in each group, A, B and C. Table \ref{tab:demograph} shows demographic
characteristics of the participants. The mean per-participant duration
of the study was 45 minutes (for six clinical cases).

\begin{figure}[t]
\begin{centering}
\includegraphics[width=1\columnwidth]{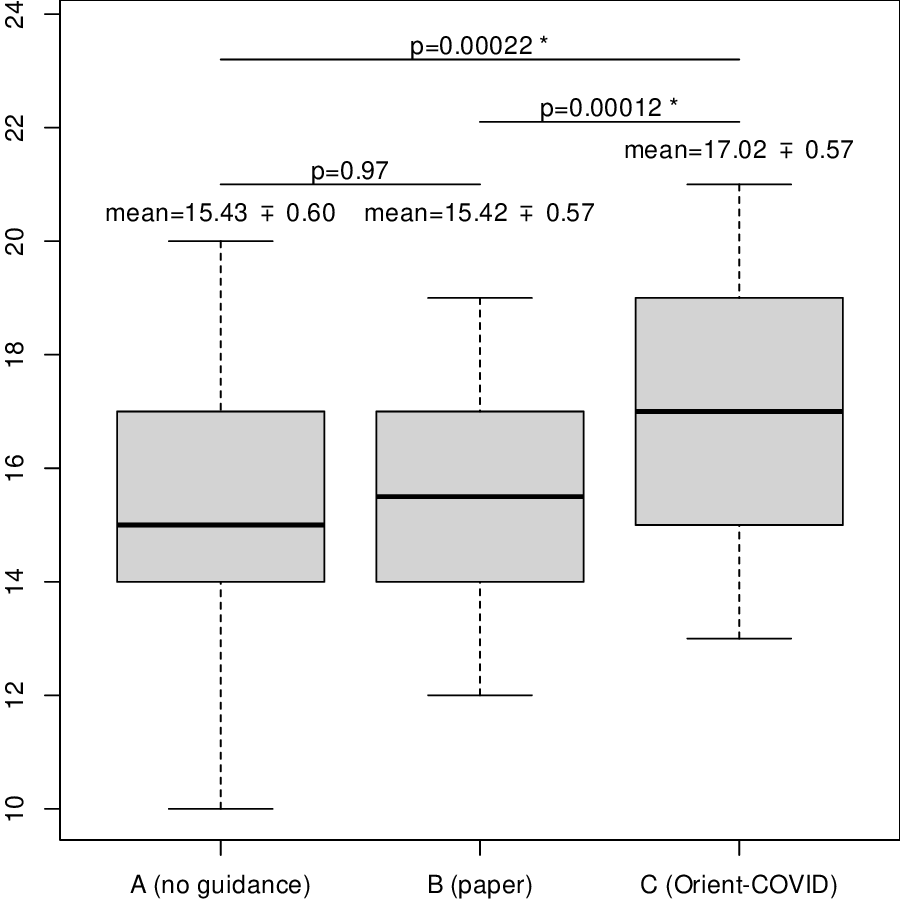}
\par\end{centering}
\caption{\label{fig:Box-plot-showing}Boxplot showing the score obtained for
each group, the mean and the 95\% confidence intervals, and \emph{p}-values
for two-by-two comparisons ({*} = significant difference).}
\end{figure}

\begin{comment}
\begin{table}[t]
\noindent \begin{centering}
\begin{tabular}{lr>{\raggedright}p{2.1cm}c}
\textbf{Group} & \textbf{Score} & \textbf{\emph{p}}\textbf{-value} & \tabularnewline
\hline 
A (no guidance) & 15.43 & \multirow{3}{2.1cm}{$\Bigr]p=0.97$\\
$\Bigr]p=0.00012${*}} & \multirow{3}{*}{\hskip-0.3cm$\Biggr]p=0.00022${*}}\tabularnewline
B (paper) & 15.42 &  & \tabularnewline
C (Orient-COVID) & 17.02 &  & \tabularnewline
\end{tabular}
\par\end{centering}
\caption{\label{tab:Mean-scores}Mean scores for each group and associated
\emph{p}-values ({*} = significant difference).}
\end{table}
\end{comment}

\subsection{Comparison of the three groups}

Figure \ref{fig:Box-plot-showing} shows the scores obtained for each
of the three groups, A (no guidance), B (paper guideline) and C (Orient-COVID).
Groups A and B obtained almost the same results, while group C obtained
a better result (about 1.6 points above). Statistical tests show that
there is no significant difference between group A and B, but a significant
difference between groups A and C, and groups B and C, respectively
($p<0.0003$).

This suggests that paper guideline did not help the participants for
solving the clinical cases, while Orient-COVID significantly improved
the quality of the decisions. The improvement is relatively modest
(1.6 points) but highly significant. This may be related to the fact
that several criteria were actually easy to answer: for 6 criteria,
the percentage of good answers is above 95\% (Chest CT, General blood
test, CRP, Decision to hospitalize, Antibiotics, Reevaluation decision).
Consequently, the difference measured is restricted to a limited number
of criteria.

As paper guidelines did not provide any support, and group A and B
performed equivalently, we grouped them in a group labeled AB (without
Orient-COVID) for the rest of the analysis, thus considering an ``Orient-COVID''
boolean variable instead of three groups. This facilitates the analysis
and increases group size.

\begin{table}[t]
\begin{centering}
\begin{tabular}{rllllc}
 & \textbf{Criteria} & \textbf{AB} & \textbf{C} & \textbf{\emph{p}}\textbf{-value} & \tabularnewline
\hline 
\multirow{9}{*}{\begin{turn}{90}
Initial evaluation
\end{turn}} & EKG & 0.53 & 0.73 & 0.0053 & .\tabularnewline
 & Chest CT & 0.98 & 0.98 & 0.70 & \tabularnewline
 & General blood test & 1 & 1 & -- & \tabularnewline
 & CRP & 0.96 & 0.95 & 0.81 & \tabularnewline
 & LDH & 0.43 & 0.45 & 0.75 & \tabularnewline
 & Troponin & 0.31 & \textbf{0.57} & 0.0011 & {*}\tabularnewline
 & D-Dimers & 0.61 & 0.72 & 0.14 & \tabularnewline
 & Ferritin & \textbf{0.25} & 0 & $5.3\times10^{-9}$ & {*}\tabularnewline
 & Il 6 & 0.48 & 0.63 & 0.044 & .\tabularnewline
\hline 
\multirow{6}{*}{\begin{turn}{90}
Initial decision
\end{turn}} & Decision to hospitalize & 0.99 & 1 & 0.32 & \tabularnewline
 & Level of care & 0.86 & 0.92 & 0.23 & \tabularnewline
 & Antibiotics & 0.98 & 0.92 & 0.14 & \tabularnewline
 & Steroids & 0.85 & 0.87 & 0.76 & \tabularnewline
 & Anticoagulant & 0.70 & \textbf{0.98} & $3.6\times10^{-9}$ & {*}\tabularnewline
 & Oxygen & 0.58 & \textbf{0.82} & 0.00074 & {*}\tabularnewline
\hline 
\multirow{7}{*}{\begin{turn}{90}
Reevaluation
\end{turn}} & Clinical status & 0.73 & \textbf{0.93} & $9.8\times10^{-5}$ & {*}\tabularnewline
 & Oxygen need & 0.73 & 0.82 & 0.16 & \tabularnewline
 & Fever & 0.18 & 0.25 & 0.32 & \tabularnewline
 & Blood test & 0.88 & 0.97 & 0.018 & .\tabularnewline
 & Chest CT & 0.98 & 0.98 & 0.705 & \tabularnewline
 & Reevaluation decision & 0.96 & 1 & 0.025 & .\tabularnewline
 & Plan of care & 0.80 & 0.82 & 0.79 & \tabularnewline
\end{tabular}
\par\end{centering}
\caption{\label{tab:Per-criteria-analysis}Per criteria analysis showing the
mean score obtained for each criterion without Orient-COVID (AB) or
with Orient-COVID (C), and the corresponding \emph{p}-value ({*} :
significant after Bonferroni correction i.e. $p<0.0023$, . : $p<0.05$).}
\end{table}

\subsection{Per-criterion analysis}

Table \ref{tab:Per-criteria-analysis} shows the per-criterion analysis.
Five significant differences were observed. For four criteria (Troponin,
Anticoagulant, Oxygen and Clinical status), the value was significantly
better with Orient-COVID. For the last criteria (Ferritin), the value
was significantly lower. Indeed, it appears that ferritin lab test
was not considered in the Orient-COVID decision support tool. Finally,
some other criteria, \emph{e.g.} EKG or Blood test, are not significantly
impacted by the use of Orient-COVID using the Bonferroni correction,
but trends can be observed.

\begin{table}[t]
\begin{centering}
\begin{comment}
\begin{center}
\begin{tabular}{lll}
\textbf{Factor} & \textbf{\emph{p}}\textbf{-value} & \textbf{Inter. }\textbf{\emph{p}}\textbf{-value}\tabularnewline
\hline 
Clinical case ID & \textbf{$\boldsymbol{<3.9\times10^{-15}}$~~{*}} & 0.21\tabularnewline
University grade & 0.10 & 0.51\tabularnewline
\#Covid-19 cases treated & \textbf{0.010~~{*}} & 0.056\tabularnewline
Age & \textbf{$\boldsymbol{<4.4\times10^{-6}}$~~{*}} & 0.76\tabularnewline
Sex & 0.74 & 0.19\tabularnewline
Use of Orient-COVID & \textbf{0.00026~~{*}} & -\tabularnewline
\end{tabular}
\par\end{center}
\end{comment}
\par\end{centering}
\begin{centering}
\begin{tabular}{lll}
\textbf{Factor} & \textbf{\emph{p}}\textbf{-value} & \textbf{Inter. }\textbf{\emph{p}}\textbf{-value}\tabularnewline
\hline 
Clinical case ID & \textbf{$\boldsymbol{<2\times10^{-16}}$~~{*}} & \textbf{0.023 ({*})}\tabularnewline
University grade & 0.57 & 0.65\tabularnewline
\#Covid-19 cases treated & 0.31 & 0.74\tabularnewline
Age & 0.24 & 0.87\tabularnewline
Sex & 0.96 & 0.94\tabularnewline
\hline 
\end{tabular}
\par\end{centering}
\caption{\label{tab:Factor-table}Results of the LMM analysis for the various
factors considered. For each factor, the \emph{p}-value is given,
as well as the \emph{p}-value of the factor interaction with the use
of Orient-COVID.}
\end{table}

\subsection{Factor analysis}

Table \ref{tab:Factor-table} shows the per-factor analysis. No significant
difference was observed with regard to sex, age, grades obtained at
university, or according to the number of Covid-19 cases the participants
encountered during their clinical activity, nor any interaction between
these factors and Orient-COVID. Nevertheless, this analysis should
be considered cautiously because of the low group size.

\begin{figure}[t]
\begin{centering}
\includegraphics[width=1\columnwidth]{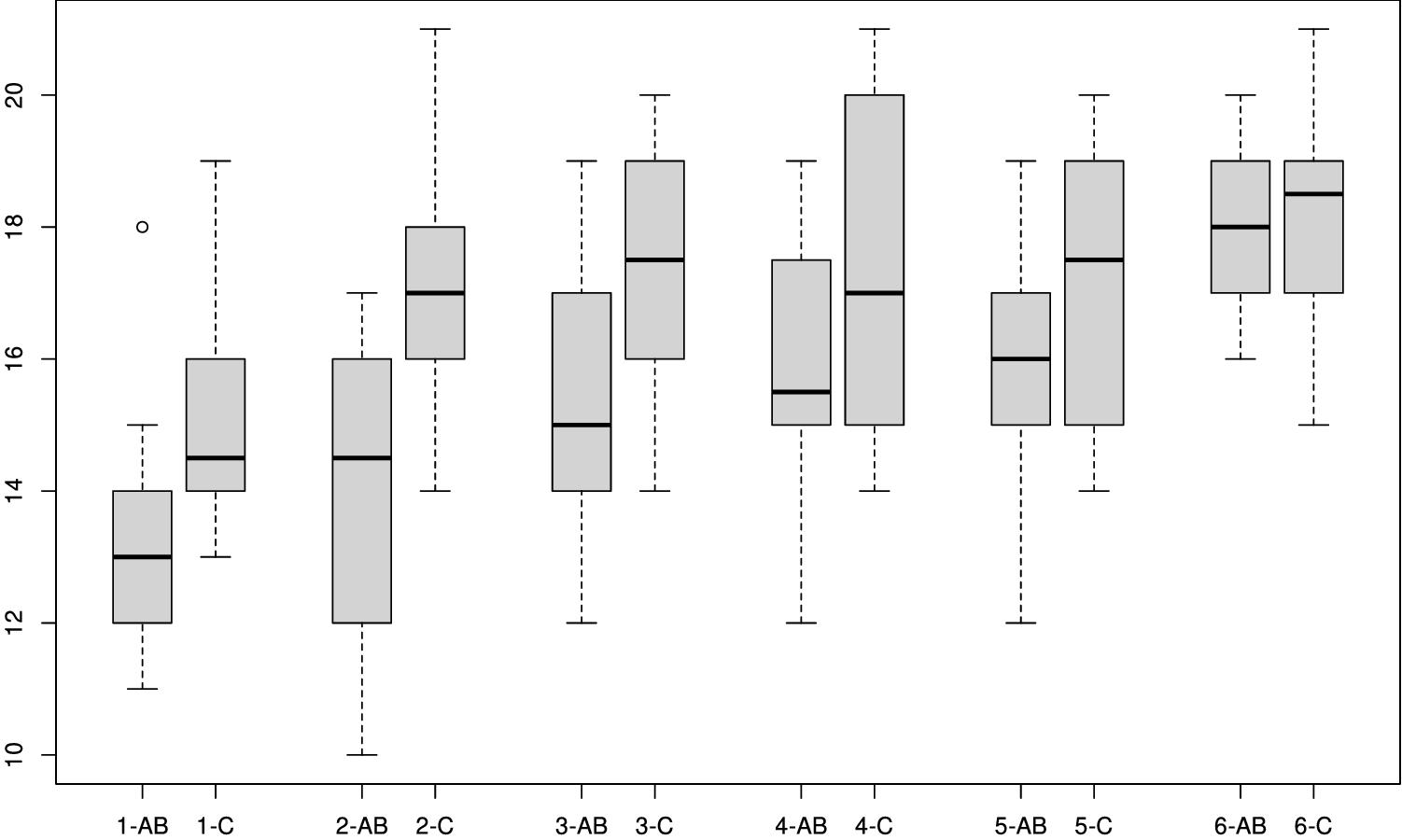}
\par\end{centering}
\caption{\label{fig:Box-plot-questions}Boxplot showing the score obtained
for each clinical case (labeled as 1-6 numbers in the box label at
the bottom), without or with Orient-COVID (labeled in the box label
as AB and C, respectively).}
\end{figure}

A significant difference was observed with clinical case ID ($p<2\times10^{-16}$),
with an interaction with Orient-COVID ($p=0.023$). Figure \ref{fig:Box-plot-questions}
shows per-case boxplots. It suggests that there was possibly some
carryover or learning effect, \emph{i.e.} the participants improved
their performance as they solved clinical cases. The use of Orient-COVID
improved the score for all clinical cases, with the exception of the
clinical case \#6.

\section{\label{sec:Discussion}Discussion}

In this study, we have assessed the impact of the use of Orient-COVID,
a computerized CDSS based on the advanced visualization of decision
trees, on the adherence to COVID19 CPGs during a randomized controlled
trial in simulated environment. The results shows a significant overall
positive impact of the use of the Orient-COVID CDSS versus paper guidelines
and absence of guidance. Previously, we performed a preliminary qualitative
evaluation of Orient-COVID perceived usability, and we obtained a
SUS (System Usability Scale) score of 92.5\% \citep{Lamy2023}, which
is ``excellent'' according to the SUS scale \citep{Bangor2009}.

\subsection{Limitations}

A study limitation is that participants were medical students. This
choice could have potentially affected the results. However, students
are very involved in decision-making, especially in university medical
centers. Another limit is that the study was monocentric. Finally,
the study is a simulation trial \citep{Cheng2016}, in which participants
may not have acted as they would on real patients; however, simulation
trials are often used for CDSSs because of their simplicity to set
up, \emph{e.g.} \citep{Higi2023}.

\begin{comment}
A second limitation is that the evaluation score was based on expert
opinion. It may introduce some biases or restrictions. However, it
attempted to grasp important decision criteria and steps across the
assessment, therapeutic decision and reassessment phases, in accordance
with international COVID19 CPGs.
\end{comment}

The main limitation of Orient-COVID is that it relies on CPGs, and
therefore on their clinical quality. However, it has been shown that
the quality of COVID-19 guidelines was not encouragingly high \citep{Wang2021,Amer2022,Burns2021}.
Another limitation is that it is designing for presenting the decision
tree in its entirety on the screen, which is feasible on a computer,
but not on a smaller screen, such as the one of a smartphone.

\subsection{Comparison to literature}

In the literature, many CDSS were proposed for Covid-19. A first review,
by A Ameri \emph{et al.} \citep{Ameri2024}, distinguished two main
approaches: (1) expert-system CDSSs that rely on a human-designed
knowledge base, such as those implementing CPGs, and (2) CDSSs that
rely on machine learning, for which the knowledge is learned from
huge patient databases. In the first category, to which belongs Orient-COVID,
the most common approach was rule-based systems. Most of the proposed
CDSSs (about 75\%) belong to the second category.

A second review, by H Ben Khalfallah \emph{et al.} \citep{Ben2023},
distinguished four categories of CDSSs: (1) alert systems that raise
alerts at the point of care, (2) monitoring systems that track and
record various physiological parameters of patients, (3) recommendation
systems that support the navigation through CPGs, and (4) prediction
systems that aim at making diagnosis or predicting the outcomes of
treatment.

\begin{comment}
An example of guideline-based CDSS for COVID-19 is the system proposed
by JP Vrel \emph{et al.} \citep{Vrel2021}. It presents to the clinicians
a form to fill with medical data, and uses a decision tree internally;
but contrary to Orient-COVID, it does not permit the navigation in
the tree. Another example is CORAL \citep{Dugdale2021}, a COvid Risk
cALculator that guides clinicians to a risk-stratified COVID-19 diagnostic.

Beyond COVID-19 management, other CDSSs proposed the interactive navigation
in a decision tree. The most common approach is to display the entire
tree in a panel, and the details of the current node in another panel,
\emph{e.g.} in the CDSS by JN Babione \emph{et al.} for pulmonary
embolism \citep{Babione2020}. 
\end{comment}

\subsection{Detailed impact on the adherence to COVID19 CPGs}

A positive impact was observed for certain criteria pertaining to
all three levels of the clinical management (initial assessment, therapeutic
decisions and reassessment), namely: Troponin, Anticoagulation treatment,
Oxygenation treatment and Clinical status, with additional trends
on the following criteria: EKG, Il 6, Blood test and Reevaluation
decision. For COVID19, EKG and troponin measurement upon admission
were reported as having a potentially high impact on the morbidity
and mortality of COVID19 patients \citep{Wibowo2021,Changal2021,Mele2022,Zeijlon2022},
and were used in prognostic scores \citep{Appel2024,Buttia2023}.
Moreover, anticoagulation use has been found to be associated with
better clinical outcomes for COVID19 patients \citep{Jiang2021}.
Finally, oxygenation supplementation is critical \citep{Mansab2021}.

Regarding the major decisions relative to patient admission, transfer
or discharge, it seems, however, that there is no significant improvement
associated with the use of Orient-COVID. This might imply that the
real added value of the CDSS might lie more in guiding the clinician
for the details of the evaluation and therapeutics, rather than result
in a change in the distribution of COVID19 patient across different
hospitals and extra-hospital settings.

Regarding ferritin, we have seen that Orient-COVID provided no support,
and that it led to a significantly lower adherence on that point,
probably because participants expected some guidance. This phenomenon
is known as automation bias \citep{Goddard2014}.

\subsection{Comparison to paper CPG experience}

The study results have shown no significant difference between paper
guidance and absence of guidance, and a significant difference between
the CDSS and both other groups. This is in line with other studies
reporting that the paper-based guidelines did not support sufficiently
healthcare practitioners in finding patient-specific recommendations
\citep{Kilsdonk2016,OConnor2016}.

Three advantages were reported orally by participants during the session:
the intuitive aspect and functionalities of the user interface, the
ease of navigation in the decision tree, and the automatic navigation
after having entered patient data. This can reduce the time to decision
and the cognitive burden \citep{Sanderson2023}.

\subsection{Challenges and perspectives of integration in real clinical workflow}

Orient-COVID was constructed using an ontological approach. It makes
its update easy, since, in case of change in the recommendations,
editing the decision trees modeled in the ontology is sufficient for
updating the system, without any need to modify the implementation.
In fact, ontologies facilitate standardization, flexibility for change,
and therefore promote sharing and reusability of medical knowledge
between CDSS systems implemented in different technologies and standards.
Along with the decision support tool, we developed a dedicated decision
tree editor as a desktop application.

Further evaluations of the approach are, of course, needed, to assess
its usability more in depth, but also to evaluate it in terms of chance
of erroneous navigation and time gain for clinicians. The semi-automatic
navigation, considering structured patient data available, also has
to be connected to EHR from hospitals to reduce the cognitive burden
associated with data entry, and properly evaluated.

\section{\label{sec:Conclusion}Conclusion}

We presented a simulation-based evaluation of Orient-COVID, a clinical
decision support system for COVID19. The results showed that this
tool significantly improved the adherence of participants to guidelines
when compared to paper-based guidance and absence of guidance. In
particular, adherence to a number of important assessment and therapeutic
criteria were significantly improved, which might translate into better
decisions impacting patient morbidity and mortality. Our main perspectives
include the integration of the system with hospital EHR, and the application
of the dynamic multi-path decision tree visual approach to other clinical
guidelines, in order to support clinicians on multiple types of patient
diagnostic or therapeutic decisions for other clinical situations
beyond COVID-19.

\section{Summary table}

\subsection{What was already known on the topic}
\begin{itemize}
\item Clinician adherence to clinical practice guidelines is low for many
disorders, including COVID-19.
\item Clinical decision support systems implementing guidelines can improve
the clinician's adherence to guidelines.
\item An approach is to permit the navigation through the guidelines, presented
as a decision tree.
\item This approach is limited by the size of the tree, which rapidly grows
and does not allow its visualization in its entirety on the screen.
\end{itemize}

\subsection{What this study added to our knowledge}
\begin{itemize}
\item Using the fisheye visualization technique and an innovative multi-path
tree model, we designed Orient-COVID, a clinical decision support
system for managing patients with COVID-19.
\item We conducted a randomized controlled trial in a near-real simulation
setting comparing Orient-COVID to paper guidelines and to the absence
of guidance.
\item Results showed that Orient-COVID improved significantly guideline
adherence compared to paper guidelines or the absence of guidance.
\end{itemize}
\begin{comment}

\section{Statement of significance}

\subsection{Problem or Issue}

Clinician adherence to clinical practice guidelines is low for many
disorders, including COVID-19.

\subsection{What is Already Known}

Clinical decision support systems implementing guidelines can improve
the clinician's adherence to guidelines. An approach is to permit
the navigation through the guidelines, presented as a decision tree.
This approach is limited by the size of the tree, which rapidly grows
and does not allow its visualization in its entirety on the screen.

\subsection{What this Paper Adds}

Using the fisheye visualization technique and an innovative multi-path
tree model, we designed a clinical decision support system for managing
patients with COVID-19. A randomized controlled trial in a near-real
simulation setting showed that this tool, named Orient-COVID, improved
significantly the guidelines ahderence, when compared with paper guidelines
or with the absence of guidance.
\end{comment}

\section*{CRediT authorship contribution statement}

Mouin Jammal: Methodology, Formal analysis, Clinical validation, Investigation,
Writing - Original Draft

Antoine Saab: Conceptualization, Methodology, Formal analysis, Investigation,
Supervision, Project administration, Funding acquisition, Writing-
Original Draft

Cynthia Abi Khalil: Conceptualization, Methodology, Clinical validation,
Investigation, Review \& Editing

Charbel Mourad: Project administration, Funding acquisition

Rosy Tsopra: Methodology, Writing - Review \& Editing

Melody Saikali: Conceptualization, Methodology, Investigation

Jean-Baptiste Lamy: Conceptualization, Methodology, Software, Formal
analysis, Visualization, Supervision, Project administration, Funding
acquisition, Writing - Original Draft

\begin{comment}
MS, AS, CAB, CM and MS selected the CPGs and developed the decision
trees.

JBL formalized them in OWL and implemented the web application. 

AS, RT and JBL set up the study protocol. 

MS, AS, CAB, CM and MS recruited the participants and organized the
study.

JBL performed the statistical analysis.

MS, AS, CAB, CM, MS and JBL drafted the initial manuscript.

All authors approved the final version of the manuscript.
\end{comment}

\section*{Declarations of interest}

None

\section*{Acknowledgments}

This work was funded by the French Research Agency (ANR) and the Lebanese
national research center (CNRS-L) through the Orient-COVID project
{[}Grant No. \mbox{ANR-21-LIBA-0004}{]}.

\bibliographystyle{elsarticle-num}
\bibliography{biblio_ama}

\end{document}